\documentclass[]{jkas} 

\def\year{2026} 
\def\volume{1} 
\def\issue{1} 
\def\beginpage{1} 
\def\received{---} 
\def\accepted{---} 
\def\published{---} 
\date{Received \received; Accepted \accepted; Published \published}

\newcommand{\refbf}{}

\title{%
High Resolution Spectroscopic Analysis of Chromospheric Line Evolution during an Energetic Flare on AD~Leo
}

\author[1]{Younghun~Oh}{0009-0004-3319-3731}
\author[1,2,$\star$]{Seo-Won~Chang}{0000-0002-3118-8275}
\author[1,$\star$]{Jongchul~Chae}{0000-0002-7073-868X}
\author[3]{Juhyung~Kang}{0000-0003-3540-4112}
\author[1]{Soosang~Kang}{0000-0002-3657-4845}
\author[1]{Kyeore~Lee}{0000-0001-9455-3615}
\author[1,2]{Kyoung-Sun~Lee}{0000-0002-4329-9546}
\author[3,4]{Eun-Kyung~Lim}{0000-0002-7358-9827}
\author[3]{Hyun-Il~Sung}{0000-0001-9515-3584}

\affil[1]{Department of Physics \& Astronomy, Seoul National University, Seoul 08826, Republic of Korea}
\affil[2]{SNU Astronomy Research Center, Astronomy Program, Department of Physics and Astronomy, Seoul National University, Gwanak-gu, Seoul 08826, Republic of Korea}
\affil[3]{Korea Astronomy and Space Science Institute, Daejeon 34055, Republic of Korea}
\affil[4]{Astronomy and Space Science, University of Science and Technology, 217 Gajeong-ro, Yuseong-gu, Daejeon 34113, Republic of Korea}

\def\corrauthor{%
S.-W. Chang, \email{seowon.chang@snu.ac.kr},
J. Chae, \email{jcchae@snu.ac.kr}
}

\def\runningauthor{%
Oh et al.
}

\def\runningtitle{%
High-Resolution Spectroscopy of AD Leo Superflare
}

\def\keywords{%
stars: flare --- stars: individual: AD~Leo --- stars: chromospheres --- techniques: spectroscopic
}

\def\abstracttext{%
{\refbf Active M dwarfs exhibit frequent and energetic flares that provide a unique laboratory for studying chromospheric heating processes under extreme magnetic activity.} To probe the flare process of M-dwarfs, we present a high-resolution ($R\sim30,000$) spectroscopic case study of a superflare on AD~Leo, detected on 2023 March 14 using the Bohyunsan Optical Echelle Spectrograph (BOES). {\refbf Such high-energy events are rarely captured with simultaneous multi-line spectroscopy, allowing us to trace the energy partition and temporal evolution of the chromospheric lines.} Based on equivalent width variations, we found that the H$\alpha$ line radiated $8.8\times10^{30}$~erg, implying a total bolometric energy ($\sim10^{33}$ erg) comparable to the largest solar flares. {\refbf The Balmer series dominated the energy budget; the individual Ca~{\sc ii} H and K lines contributed 47.5\% and 26.2\% of the H$\alpha$ energy, respectively, while each Ca~{\sc ii} infrared triplet line emitted $\sim17$--19\%.} We confirm that the delayed peak emission, previously reported for Ca~{\sc ii} H\&K, also occurs in the Ca~{\sc ii} triplet and Na~{\sc i} lines. These delays are consistent with the Neupert effect, suggesting that cumulative heating governs the gradual phase emission. While this superflare resembles solar flares in general morphology, {\refbf it also displayed systematic differences in chromospheric emission. It is likely that these differences reflect the distinct atmospheric structure and quiescent chromospheric conditions of M dwarfs, rather than fundamentally different flare physics.}
}

\begin{document}
\jkashead 


\section{Introduction\label{sec:intro}}
Stellar flares are energetic and transient events characterized by a sudden release of magnetic energy through magnetic reconnection in stellar atmospheres. In the standard flare model, often referred to as the CSHKP (Carmichael-Sturrock-Hirayama-Kopp-Pneuman) model, this released energy accelerates particles and heats the plasma, leading to chromospheric evaporation and subsequent emission across the electromagnetic spectrum \citep{1964NASSP..50..451C,1966ApJ...143....3S,1974SoPh...34..323H,1976SoPh...50...85K}. While solar flares have been extensively investigated with spatially resolved observations, stellar flares remain poorly understood due to the lack of spatial resolution and limited spectroscopic coverage. As a result, our understanding of stellar flares relies heavily on analogies with solar phenomena and modeling techniques calibrated on solar data \citep{2024LRSP...21....1K}. 

{\refbf Stellar flares are observed across a wide range of spectral types, from cool K--M dwarfs to hot A-type stars \citep{2015MNRAS.447.2714B}. Among these, M dwarfs} are known to produce frequent and powerful flares, sometimes exceeding $10^{33}$ erg in bolometric energy--so-called superflares {\refbf (e.g., \citealt{2020AJ....159...60G})}. These energetic outbursts are driven by strong magnetic activity, which is sustained in late-type M dwarfs (typically later than M3--M4) by their fully (or nearly fully) convective interiors and rapid rotation {\refbf \citep{2016ApJ...829...23D,2019ApJS..241...29Y}}. {\refbf Such flares, together with associated coronal mass ejections (CMEs) and energetic particle events, are collectively referred to as stellar storms in analogy with solar storms.} From an exoplanetary perspective, such intense activity poses significant challenges to habitability. Enhanced fluxes of X-ray and extreme ultraviolet (XUV) radiation{\refbf , along with potential CMEs and energetic particles,} can profoundly affect the atmospheric evolution of close-in exoplanets {\refbf (e.g., \citealt{2007AsBio...7..167K,2007AsBio...7..185L,2016ApJ...826..195K})}. Therefore, characterizing the energy budget and radiation properties of these flares is crucial for assessing the environments of planets orbiting active stars.

Although photometric missions such as \textit{Kepler} and the Transiting Exoplanet Survey Satellite (\textit{TESS}) have provided vast amounts of data on the statistical properties and frequency of flares on M dwarfs (e.g., \citealt{2016ApJ...829...23D,2020AJ....159...60G, 2021A&A...645A..42I,2022ApJ...935..143P}), light curves alone are insufficient to diagnose the physical conditions of the flaring plasma. High-resolution spectroscopy is essential for probing key physical parameters such as plasma heating, line broadening, and detailed energy distribution across different chromospheric layers. Furthermore, time-resolved spectroscopy allows us to investigate the temporal evolution of chromospheric lines, providing insights into heating mechanisms such as the Neupert effect \citep{1968ApJ...153L..59N}, where the derivative of the soft X-ray (or gradual phase) emission correlates with the impulsive hard X-ray emission. 

{\refbf The Neupert effect has been observed in most large solar flares \citep{1993SoPh..146..177D} and was subsequently confirmed in M~dwarf flares through multiwavelength campaigns combining optical and soft X-ray or radio observations (e.g., \citealt{1995ApJ...453..464H, 1996ApJ...471.1002G}). However, verifying the effect in stellar flares remains challenging, as it requires simultaneous coverage across multiple wavelength regimes with sufficient time resolution to separate the impulsive and gradual phases. High-resolution spectroscopy enables such studies to resolve the temporal responses of individual chromospheric lines that are inaccessible from broadband photometry alone \citep{2011A&A...534A.133F}.}

Despite extensive photometric monitoring of stellar flares, simultaneous multi-line spectroscopic observations remain rare. AD Leonis (AD Leo; Gliese 388) has served as a benchmark target for such studies due to its high activity. \citet{2003ApJ...597..535H} utilized simultaneous high-resolution optical ($R\sim55,000$; 3800--8800~\AA) and UV ($R\sim70,000$; 1160--1700~\AA) spectroscopy to identify common heating mechanisms, yet the total radiated energies of these events were limited to the range of $10^{31}$--$10^{32}$ erg. More recently, \citet{2020A&A...637A..13M} and \citet{2024ApJ...961..189N} utilized high spectral resolution ($R\sim$~30,000--35,000) to search for chromospheric line asymmetries and potential mass ejections. However, their analyses focused primarily on kinematic signatures rather than the detailed temporal evolution and energy partition of the chromospheric lines during a superflare. In contrast, \citet{2013ApJS..207...15K} utilized broad wavelength coverage (3400--9200~\AA) to investigate the temporal evolution of hydrogen Balmer lines and continuum components. However, their study utilized low-resolution spectroscopy ($R\sim$ 600--800), which limited the analysis of detailed line profiles. 

This study builds on the analysis first presented in the Master’s thesis of \citet{Oh2025}, {\refbf which presented the initial spectroscopic analysis of the AD Leo superflare observed on 2023 March 14. Relative to that work, we improve the line-energy measurement by integrating the full time-resolved $\Delta$EW light curves, add a discussion of the Ca~{\sc ii} H\&K line evolution, and extend the comparison with previous AD~Leo and solar flare
studies.} By analyzing the equivalent width (EW) variations of multiple chromospheric diagnostics—including the hydrogen Balmer series, He~{\sc i}, and {\refbf metallic lines (e.g., Ca~{\sc ii} and Na~{\sc i})}—we aim to quantify the radiated energy budget and investigate the detailed atmospheric response. The paper is organized as follows: Section~\ref{sec:obsdata} describes the observations and data reduction. Section~\ref{sec:analysis} outlines the analysis methods. Section~\ref{sec:results} presents the flare morphology, radiated energy budget, and temporal evolution. Section~\ref{sec:discussion} discusses the physical implications, focusing on the Neupert effect and a comparison with solar flares. Finally, Section~\ref{sec:summary} summarizes our findings.

\begin{table}[t!]
    \caption{Basic properties of AD~Leo.\label{tab:adleo}}
    \centering
    \begin{tabular}{lcc}
        \toprule
        Parameter & Value & Ref. \\
        \midrule
        Spectral Type         & dM3.5Ve        & (1) \\
        V mag           & 9.52             & (2) \\
        Distance              & 4.96 pc          & (3) \\
        Mass                  & $0.42~M_{\odot}$  & (4) \\        
        Radius                & $0.39~R_{\odot}$ & (4) \\
        Luminosity            & $0.02~L_{\odot}$ & (4) \\
        Effective Temperature & 3477 K           & (5) \\
        <B>       & 2.8--3.6~kG   & (6) \\
        Rotation Period       & 2.23 days        & (7) \\
        Age                   & 100--300 Myr      & (1,6) \\
        \bottomrule
    \end{tabular}
    \begin{flushleft}
        \footnotesize
        References. 
        (1) \citet{2009ApJ...699..649S}; 
        (2) \citet{2013AJ....145...44Z}; 
        (3) \citet{2021A&A...649A...1G}; 
        (4) \citet{2023MNRAS.522.1342C};
        (5) \citet{2022A&A...666A.143K}; 
        (6) \citet{2023A&A...676A..56B};
        (7) \citet{2008MNRAS.390..567M}.
    \end{flushleft}
\end{table} 

\section{Observations and Data Reduction} \label{sec:obsdata}

\subsection{Target Description and Observations}

AD~Leo is a prototypical active M-dwarf located in the immediate solar neighborhood ($d \approx 5$~pc; \citealt{2021A&A...649A...1G}) and is a member of the Castor Moving Group \citep{2009ApJ...699..649S}. With a brightness of $V \approx 9.5$~mag, it is one of the brightest M-dwarfs accessible to high-resolution spectroscopy and is renowned for its intense magnetic activity. {\refbf While the star exhibits frequent flaring, the reported rates depend on the energy range and detection sensitivity. Ground-based spectroscopic monitoring has yielded rates of up to $\sim$17~flares~day$^{-1}$ for weak flares with estimated total energies of $\sim$10$^{30}$~erg \citep{2006A&A...452..987C}. Recent \textit{TESS} observations detected 96 flares with bolometric energies ranging from $\sim$10$^{32}$ to $\sim$10$^{35}$~erg at an overall rate of 4.65~flares~day$^{-1}$, while flares exceeding 10$^{34}$~erg occur at a rate of only $\sim$0.34~day$^{-1}$ \citep{2025ApJ...980..196R}.} Consistent with its fully convective interior, AD~Leo exhibits strong chromospheric emission and possesses a mean magnetic flux density of 2.8--3.6~kG \citep{2023A&A...676A..56B}. The star rotates rapidly with a period of 2.23 days, about twelve times faster than the Sun, which likely drives its high flare productivity and the formation of large starspots {\refbf (e.g., \citealt{2020ApJ...905..107M})}. Its rotation axis is viewed nearly pole-on \citep{2008MNRAS.390..567M,2022A&A...666A.143K}, providing a favorable geometry for monitoring flare-induced variability. A summary of the physical properties of AD~Leo is presented in Table~\ref{tab:adleo}.

We conducted high-resolution spectroscopic monitoring of AD~Leo using the Bohyunsan Optical Echelle Spectrograph (BOES; \citealt{Kim2000BOAO}) mounted on the 1.8-m telescope at the Bohyunsan Optical Astronomy Observatory (BOAO). The observing campaign spanned three nights in 2023: February 24 (12:03--12:27 UT), March 14 (11:32--14:03 UT), and December 29 (15:16--20:56 UT). Among these, the data obtained on March 14 captured the complete temporal evolution—covering both the rise and decay phases—of a superflare event, which is the primary focus of this study. Observations on this night were performed under clear sky conditions.

BOES is a fiber-fed echelle spectrograph equipped with a $2048 \times 4096$ pixel CCD, covering a wide wavelength range from 3,500~\AA{}  to 10,500~\AA{} with resolving powers between $R\sim30{,}000$ and $R\sim90{,}000$ \citep{Kim2000BOAO}. For this study, we used the low-resolution fiber ($R\sim30{,}000$) to maximize throughput, yielding an effective resolution of $\sim$~20{,}000 over the chromospheric line region of interest (3900--8700~\AA{}). The CCD has a gain of 1.9 $e^-$/ADU and a readout noise of 5.8 $e^-$. The resulting spectra provided a signal-to-noise ratio (SNR) greater than 10 over this wavelength range, enabling reliable measurements of line-profile variations during the flare events.

To resolve the temporal evolution of the flare, we performed intensive time-resolved spectroscopy, obtaining a sequence of 13 spectra over a continuous 2.5-hour window. Each frame was exposed for 600 seconds. Considering the CCD readout time of 97 seconds, the effective observational cadence was about 12 minutes. Although the number of frames is limited, this dataset successfully spans the entire duration of the event, providing sufficient temporal resolution to characterize both the impulsive rise and the gradual decay phases of the superflare. A detailed log of the observations is provided in Table~\ref{tab:log_mar14}.

\begin{table}[t!]
    \centering
    \caption{Observation Log for 2023 March 14\label{tab:log_mar14}}
    \begin{tabular}{cccc}
        \toprule
        Frame No. & Object & Exptime (s) & Exp. Start (UT) \\
        \midrule
        018303 & Bias & 0.0 & 09:56:50.48 \\
        018304 & Bias & 0.0 & 09:58:27.52 \\
        018305 & Bias & 0.0 & 10:00:04.54 \\
        018306 & Bias & 0.0 & 10:01:41.57 \\
        018307 & Bias & 0.0 & 10:03:18.59 \\
        018308 & ThAr & 0.3 & 10:05:31.06 \\
        018309 & ThAr & 1.5 & 10:07:25.97 \\
        \midrule
        018310 & Flat & 40.0 & 10:19:34.16 \\
        018311 & Flat & 40.0 & 10:21:51.32 \\
        018312 & Flat & 40.0 & 10:24:08.46 \\
        018313 & Flat & 40.0 & 10:26:25.59 \\
        018314 & Flat & 40.0 & 10:28:42.72 \\
        \midrule
        018319 & AD Leo & 600.0 & 11:32:39.39 \\
        018320 & AD Leo & 600.0 & 11:44:16.51 \\
        018321 & AD Leo & 600.0 & 11:55:53.65 \\
        018322 & AD Leo & 600.0 & 12:07:30.77 \\
        018323 & AD Leo & 600.0 & 12:19:07.89 \\
        018324 & AD Leo & 600.0 & 12:30:45.01 \\
        018325 & AD Leo & 600.0 & 12:42:22.13 \\
        018326 & AD Leo & 600.0 & 12:53:59.27 \\
        018327 & AD Leo & 600.0 & 13:05:36.40 \\
        018328 & AD Leo & 600.0 & 13:17:13.53 \\
        018329 & AD Leo & 600.0 & 13:28:50.66 \\
        018330 & AD Leo & 600.0 & 13:40:27.81 \\
        018331 & AD Leo & 600.0 & 13:52:04.97 \\
        \bottomrule
    \end{tabular}
\end{table}

\begin{table*}[t!]
    \centering
    \small
    \caption{Measured properties of the chromospheric emission lines for the 2023 March 14 superflare. $\sigma(\Delta\mathrm{EW_{max}})$ and SNR are the measurement uncertainty and the local-continuum signal-to-noise ratio at the epoch of the peak excess equivalent width, $\Delta\mathrm{EW_{max}}$. The radiated energy $\Delta E$ is the time integral of the excess line luminosity, and $t_{1/2}$ is the full width at half maximum of the $\Delta$EW light curve. \label{tab:lineparameters}}
    \begin{tabular}{lccccccc}
        \toprule
        Line & Integration Range & $F_{c}^{q}$ & $\Delta\mathrm{EW_{max}}$ & $\sigma(\Delta\mathrm{EW_{max}})$ & SNR & $\Delta E$ & $t_{1/2}$ \\ 
        (\AA) & ($\pm$\,km\,s$^{-1}$) & ($10^{-12}$\,erg\,s$^{-1}$\,cm$^{-2}$\,\AA$^{-1}$) & (\AA) & (\AA) & & ($10^{29}$\,erg) & (min) \\
        \midrule
        3934 Ca~\textsc{ii}~K   & 30  & 0.12 & 1.30 & 0.911 & 10.7  & 23.07 & 92.8 \\
        3968 Ca~\textsc{ii}~H   & 30  & 0.15 & 1.58 & 0.544 & 12.5  & 41.82 & $>$86.3 \\
        3970 H$\epsilon$        & 80  & 0.15 & 1.71 & 0.377 & 16.6  & 23.70 & \\
        4102 H$\delta$          & 120 & 0.20 & 2.02 & 0.256 & 32.7  & 45.53 & 62.5 \\
        4340 H$\gamma$          & 120 & 0.25 & 2.92 & 0.281 & 39.6  & 76.33 & 58.1 \\
        4471 He~\textsc{i}      & 30  & 0.37 & 0.12 & 0.019 & 74.0  & 4.40  & 55.3 \\
        4861 H$\beta$           & 120 & 0.49 & 1.40 & 0.069 & 153.4 & 67.84 & 53.8 \\
        5167 Mg~\textsc{i}      & 30  & 0.60 & 0.03 & 0.007 & 144.2 & 1.33  & 35.2 \\
        5169 Fe~\textsc{ii}     & 30  & 0.52 & 0.07 & 0.010 & 113.5 & 2.45  & 38.3 \\
        5173 Mg~\textsc{i}      & 30  & 0.60 & 0.04 & 0.007 & 149.8 & 1.97  & 58.5 \\
        5184 Mg~\textsc{i}      & 30  & 0.61 & 0.05 & 0.007 & 144.6 & 3.49  & 73.5 \\
        5876 He~\textsc{i}~D3   & 30  & 0.64 & 0.20 & 0.010 & 225.3 & 12.73 & 54.8 \\
        5890 Na~\textsc{i}~D2   & 25  & 0.83 & 0.08 & 0.006 & 144.1 & 7.68  & 61.0 \\
        5896 Na~\textsc{i}~D1   & 25  & 0.83 & 0.09 & 0.005 & 166.9 & 8.50  & 57.0 \\
        6563 H$\alpha$          & 120 & 1.46 & 0.52 & 0.043 & 263.0 & 88.03 & 61.8 \\
        6678 He~\textsc{i}      & 30  & 1.11 & 0.03 & 0.006 & 322.2 & 2.60  & 33.2 \\
        7065 He~\textsc{i}      & 20  & 1.51 & 0.02 & 0.002 & 650.5 & 2.90  & $>$55.4 \\
        8498 Ca~\textsc{ii}     & 20  & 2.47 & 0.06 & 0.006 & 248.0 & 16.27 & 58.6 \\
        8542 Ca~\textsc{ii}     & 20  & 2.52 & 0.06 & 0.005 & 279.5 & 15.72 & 58.8 \\
        8662 Ca~\textsc{ii}     & 20  & 2.60 & 0.05 & 0.006 & 221.0 & 15.19 & 59.9 \\
        \bottomrule
    \end{tabular}
\end{table*}

\subsection{Data Reduction}
The raw two-dimensional spectral images were reduced using standard Python-based algorithms. The reduction procedure included bias subtraction, flat-field correction, cosmic-ray removal, extraction of one-dimensional spectra, wavelength calibration, and continuum normalization.

First, master bias and flat frames were created by combining multiple calibration frames and applied to the science images. Cosmic rays were identified and removed from the 2D images using the Laplacian edge detection algorithm proposed by \citet{2001PASP..113.1420V}. 

Following the preprocessing, we performed order extraction to convert the two-dimensional echelle frames into one-dimensional spectra. In the raw frames, the flux peaks of individual echelle orders appear as curved traces along the cross-dispersion direction. For each column along the dispersion axis, we identified local maxima corresponding to the centers of the orders and defined order boundaries at the midpoints between adjacent maxima, which approximate the local minima between neighboring orders. The flux within these boundaries was then integrated to produce the one-dimensional spectrum for each order. This extraction procedure was repeated across the full dispersion axis by tracing the curvature of the orders.

Wavelength calibration was derived from ThAr arc lamp spectra acquired at the beginning of the night. We identified ThAr emission lines and matched them to laboratory reference wavelengths to determine the pixel-to-wavelength dispersion relation, which was then applied to the stellar spectra. To define the reference line centers, we adopted the systemic radial velocity of {\refbf $\sim$12.43~$\mathrm{km\,s^{-1}}$ for AD Leo \citep{2018MNRAS.475.1960F}}. Residual offsets, including those from Earth's motion and instrumental effects, were identified by inspecting prominent stellar absorption lines. These offsets were subsequently corrected by applying small wavelength shifts of $\le 0.3$~\AA{}. This procedure ensured that all spectra were aligned with the stellar rest frame.

Finally, each spectrum was normalized to its local continuum. Since simultaneous observations of a spectrophotometric standard star were not performed, absolute flux calibration was not feasible. However, for the purpose of tracking flare-induced line variations, absolute calibration is not strictly required provided the local continuum can be well defined. We treated the continuum as locally constant over a narrow spectral window of approximately 10~\AA{} centered on each chromospheric line. The continuum level was defined as the median flux within this window and normalized to unity.

\section{Data Analysis} \label{sec:analysis}
\subsection{Line Flux Measurement and Energy Estimation}\label{sec:analysis1}

To quantify the flare energy from the normalized spectra, we employed the equivalent width method. This approach is advantageous as it effectively traces relative temporal variations while mitigating errors associated with absolute continuum placement. The equivalent width is defined as:
\begin{equation} \label{eqn:EW}
    \begin{split}
        \mathrm{EW} = 
        \int_{\lambda_{1}}^{\lambda_{2}} \left( 1 - \frac{F_\lambda}{F_c} \right) d\lambda,
    \end{split}
\end{equation}
where $F_\lambda$ is the observed flux density, $F_c$ is the local continuum level (normalized to unity), and the integration limits [$\lambda_{1}, \lambda_{2}$] cover the full line profile. Under this convention, absorption lines yield positive values, whereas emission lines yield negative values. Since the chromospheric lines of AD Leo appear in emission, their measured EWs are negative.

Practical integration limits were determined empirically via visual inspection to capture the full broadened profiles while avoiding spectral contamination. The adopted wavelength ranges for each line are listed in Table~\ref{tab:lineparameters}. Prior to measurement, the spectra were smoothed using a Savitzky--Golay filter \citep{1964AnaCh..36.1627S} with a window size of 15 pixels ($\sim 0.5$~\AA) to suppress pixel-level noise while preserving the intrinsic line profiles. 

{\refbf The quiescent state is initially defined by the first AD~Leo spectrum of the observing sequence (Frame~018319; see Table~\ref{tab:log_mar14}), obtained immediately before the flare onset. For most lines this first frame indeed corresponds to the lowest pre-flare $\Delta$EW and is adopted as the reference. For a few lines, however, the minimum occurs in the second or third frame (see \ref{sec:lineprofiles}), and the reference is therefore set on a line-by-line basis, as detailed below.} 

The flare-induced excess equivalent width, $\Delta\mathrm{EW}$, is defined as the difference between the quiescent EW ($\mathrm{EW}^{q}$) and the flare EW ($\mathrm{EW}^{f}$):
\begin{equation} \label{eqn:DEW}
    \begin{split}
        \Delta\mathrm{EW} \equiv \mathrm{EW}^q-\mathrm{EW}^f=\int_{\lambda_{1}}^{\lambda_{2}} \left( -\frac{F^q_\lambda}{F^q_c} + \frac{F^f_\lambda}{F^f_c} \right) d\lambda \ \mathrm{.}
    \end{split}
\end{equation}
By this convention, $\Delta\mathrm{EW}>0$ corresponds to excess emission. Given that all spectra were normalized to the local continuum, $\Delta\mathrm{EW}$ serves as a direct proxy for changes in the line flux.

We estimated the measurement uncertainties using the error-propagation formalism of \citet{2006AN....327..862V}, which accounts for photon noise in both the line profile and the adjacent continuum. For continuum-normalized spectra, the uncertainty is given by:
\begin{equation}\label{eqn:EWerr}
\sigma(\mathrm{EW}) = \sqrt{1+\frac{\bar{F}_c}{\bar{F}}}\,
\frac{(\Delta\lambda-\mathrm{EW})}{SNR} \ \mathrm{,}
\end{equation}
where $\Delta\lambda$ is the integration width ($\lambda_2-\lambda_1$), $\bar{F}$ is the mean flux within the integration window, $\bar{F}_c$ is the continuum level (unity for normalized spectra), and $SNR$ is the signal-to-noise ratio of the local continuum. We adopted these derived $\sigma(\mathrm{EW})$ values as the uncertainties for $\Delta\mathrm{EW}$ in our time-series analysis.

The physical quantity of interest is the excess line flux ($\Delta F_\mathrm{line}$). To evaluate this from the normalized spectra, we assume that the absolute continuum level does not change significantly between the quiescent and flare states within the narrow integration window (i.e., $F^q_c \approx F^f_c$). Although continuum enhancement is a known feature of stellar flares, the optical flare continuum is spread over a broad wavelength range ($\sim1000$~\AA{}). In contrast, the line emission is confined to a very narrow interval ($\sim1$~\AA{}). Therefore, within the restricted integration limits used for $\Delta\mathrm{EW}$, the contribution of continuum enhancement is diluted and can be considered negligible to the first order relative to the strong line emission \citep[see][]{2015ApJ...809...79O}. Under this assumption, the excess line flux can be approximated as:
\begin{equation} \label{eqn:DFobs}
    \Delta F_{\mathrm{line}} \approx F^q_c \cdot \Delta\mathrm{EW}.
\end{equation}
We note that if significant continuum enhancement occurs, our derived values represent a lower limit to the true emitted energy.

\begin{figure*}[!t]
\centering
\includegraphics[width=160mm]{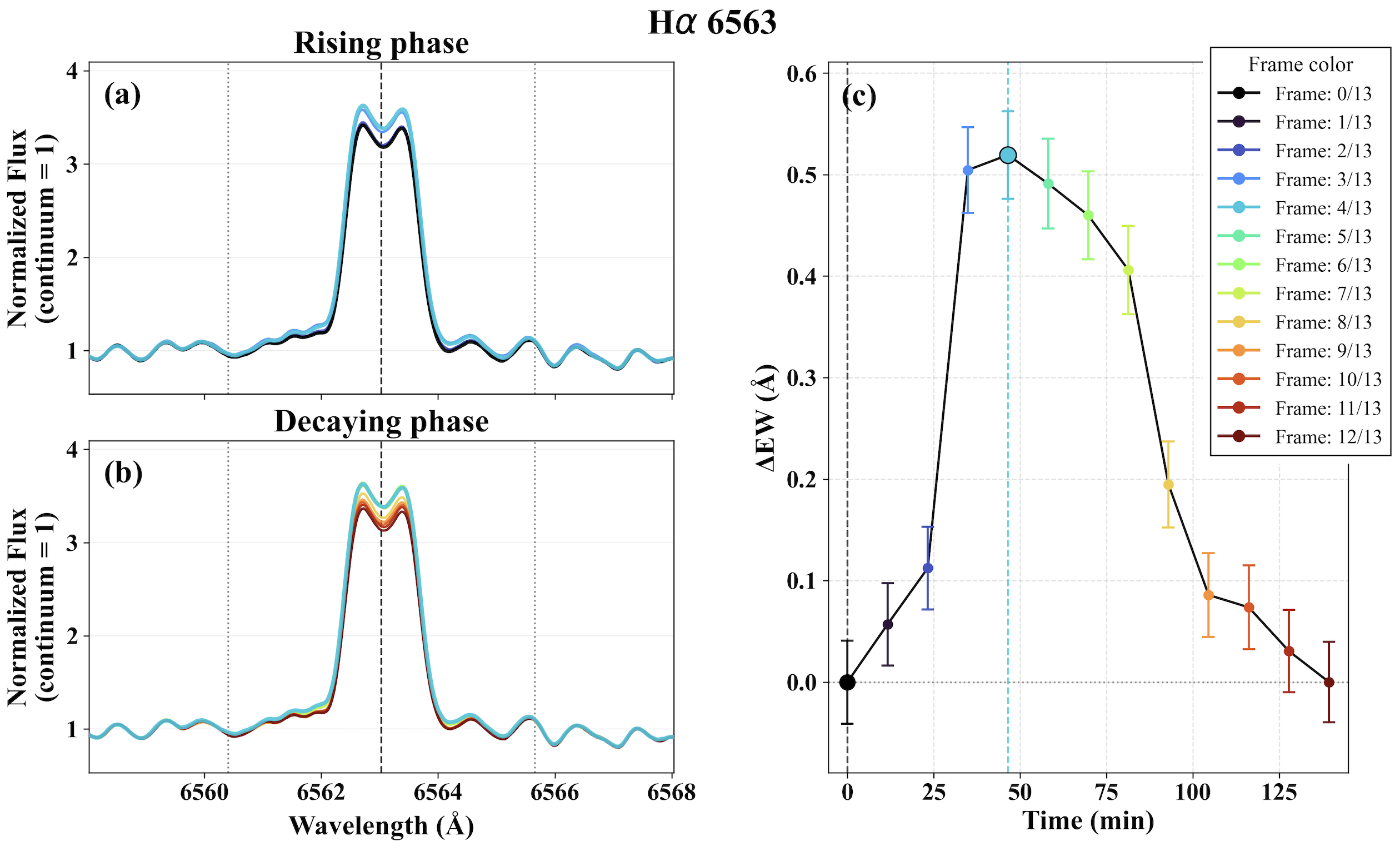} 
\caption{\refbf{Evolution of the H$\alpha$ line profile during the 2023 March 14 flare. Panels (a) and (b) show the line profiles during the rise and decay phases, respectively, with time increasing from violet to red. Vertical dashed lines mark the line center (black) and the $\pm120$~km s$^{-1}$ integration boundaries (gray). Panel (c) shows the corresponding temporal evolution of $\Delta$EW, with error bars color-coded to match the spectral frames.
\label{fig:HA1}}
}
\end{figure*}

\begin{figure*}[!t]
\centering
\includegraphics[width=160mm]{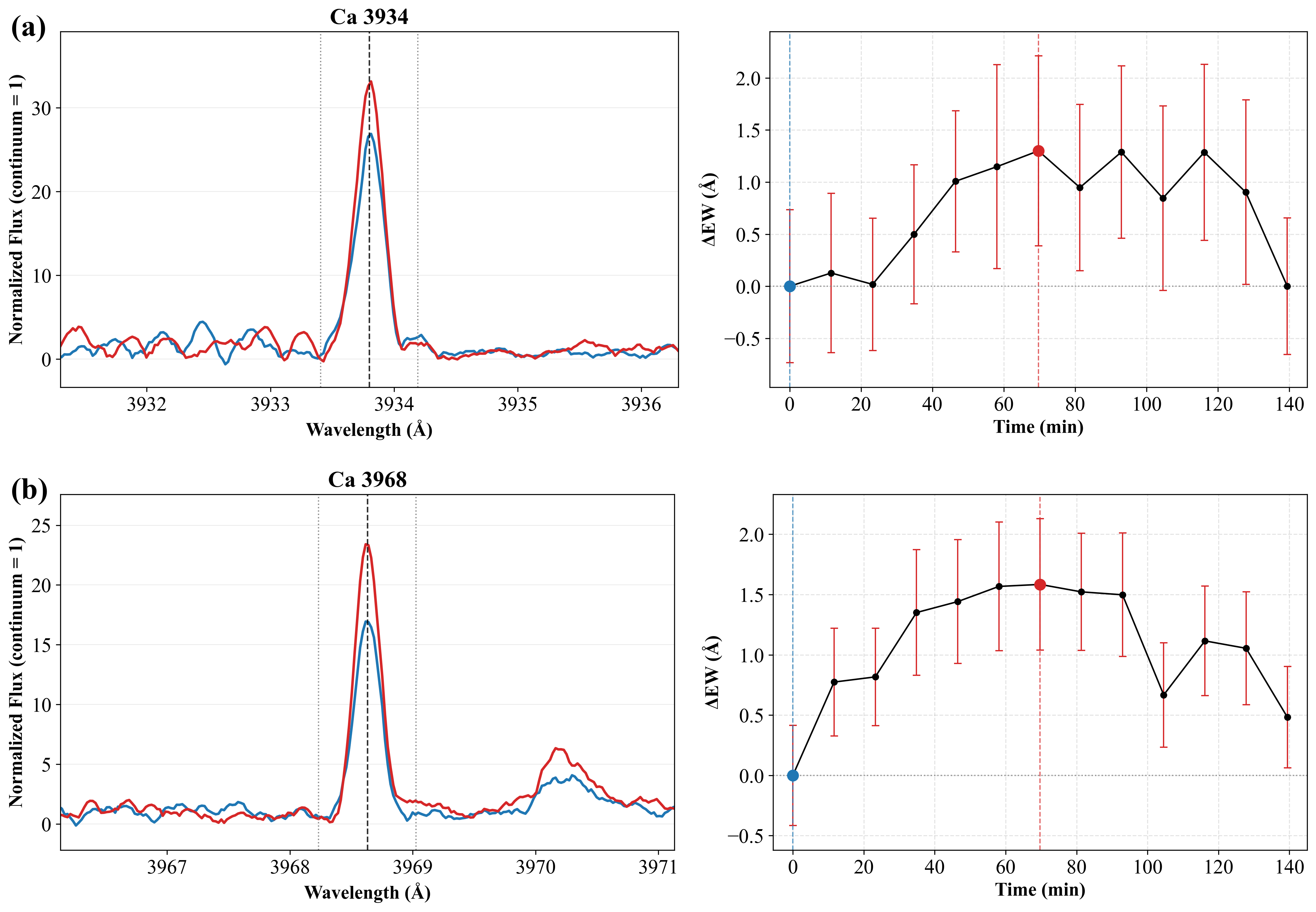}
\caption{\refbf{Time-resolved evolution of the Ca~{\sc ii} H\&K resonance lines: (a) Ca~{\sc ii} K 3934~\AA\ and (b) Ca~{\sc ii} H 3968~\AA. Left panels compare the quiescent profile (blue) with the profile at peak emission (red). The black dashed vertical line marks the line center, and the gray dotted vertical lines indicate the integration range used for the EW measurements. Right panels show the corresponding strict $\Delta\mathrm{EW}$ light curves with error bars, where the reference and peak-emission epochs are marked by blue and red vertical dashed lines, respectively.
}}
\label{fig:CA1}
\end{figure*}

To convert $\Delta\mathrm{EW}$ into physical units, the absolute flux density of the quiescent continuum ($F^q_c$) is required. In the absence of simultaneous spectrophotometric observations, we adopted the calibrated quiescent continuum flux values for AD~Leo reported by \citet{2013ApJS..207...15K}. These reference values, listed in Table~\ref{tab:lineparameters}, were derived from median fluxes in spectral windows free of strong emission lines.

The excess luminosity $\Delta L_{\mathrm{line}}$ is given by:
\begin{equation} \label{eqn:DL}
    \Delta L_{\mathrm{line}}(t) = 4\pi d^2 \Delta F_{\mathrm{line}}(t) \approx 4\pi d^2 F^q_c \cdot \Delta\mathrm{EW}(t),
\end{equation}
where $d$ is the distance to AD~Leo ($d=4.96$ pc; \citealt{2021A&A...649A...1G}). The total radiated energy was estimated by directly integrating the excess luminosity over the observed time series. This approach incorporates the time-resolved evolution of the flare, including its generally asymmetric profile with a rapid rise followed by a more gradual decay.

{\refbf Because the local continuum normalization can introduce a small offset in the quiescent $\Delta$EW values, the reference level is set for each line individually to the frame with the minimum $\Delta$EW among the first four frames, which most closely represent the quiescent stage. The excess equivalent width is then redefined so that this reference frame has $\Delta\mathrm{EW}=0$. Any negative values are set to zero, so that only the positive flare-induced excess contributes to the energy:
\begin{equation} \label{eqn:DEWpos}
    \Delta\mathrm{EW}^{+}(t_i) \equiv \max\left[\Delta\mathrm{EW}(t_i),\,0\right].
\end{equation} 

The total radiated energy for each chromospheric line was then obtained by integrating the excess luminosity over time:
\begin{equation} \label{eqn:DE}
    \Delta E = \int \Delta L_{\mathrm{line}}(t)\,dt \approx 4\pi d^2 F^q_c \int \Delta\mathrm{EW}^{+}(t)\,dt.
\end{equation}
Since the spectra were obtained at discrete observing times, we evaluated this integral using the trapezoidal rule:
\begin{equation} \label{eqn:DEtrap}
    \Delta E \approx 4\pi d^2 F^q_c
    \sum_{i=1}^{N-1}
    \frac{\Delta\mathrm{EW}^{+}_{i-1}+\Delta\mathrm{EW}^{+}_{i}}{2}
    \left(t_i-t_{i-1}\right),
\end{equation}
where $\Delta\mathrm{EW}^{+}_{i}$ is the non-negative excess equivalent width measured at time $t_i$, and $N$ is the number of spectra. This estimate is derived directly from the observed light curve and therefore reflects the measured temporal evolution of the flare. Because it is nonetheless limited by the finite observing cadence and by the assumed negligible continuum enhancement within the line window, the resulting  energies represent lower limits to the true radiated values.}

\subsection{Line Selection}
{\refbf We selected chromospheric emission lines based on two criteria: (i) lines well established as flare-sensitive diagnostics in previous stellar flare studies \citep[e.g.,][]{2005A&A...439.1137F,2006A&A...452..987C,2013ApJS..207...15K}, and (ii) lines that exhibited significant EW enhancement during the flare relative to the quiescent state in our spectra.} We applied this analysis to a set of chromospheric emission lines sensitive to flare activity covering the wavelength range from 3900 to 8700~\AA. The selected lines, listed in Table~\ref{tab:lineparameters}, include the hydrogen Balmer series (H$\alpha$ through H$\epsilon$), the ionized calcium Ca~\textsc{ii} H\&K and infrared triplet, as well as lines from neutral helium (He~\textsc{i}), magnesium (Mg~\textsc{i}), sodium (Na~\textsc{i}), and ionized iron (Fe~\textsc{ii}). {\refbf This selection provides diagnostics spanning a range of chromospheric heights. However, the number of analyzed lines is weighted toward wavelengths longer than 4000~\AA, as the blue end of our echelle spectra has lower SNR. Furthermore, our spectral coverage does not extend to the near-ultraviolet, where numerous additional chromospheric emission lines are known to exist \citep{2005A&A...439.1137F} and may contribute energy comparable to, or even greater than, that of the optical lines during the impulsive phase \citep{2003ApJ...597..535H}. The energy budget derived from the lines analyzed here should therefore be regarded as a lower limit to the total chromospheric line losses.}

\section{Results}
\label{sec:results}

\subsection{Development of the Superflare}
{\refbf Among the events detected during our campaign, the flare on 2023 March 14 stands out for its exceptional energy. With a bolometric energy of $\sim10^{33}$~erg (Section~\ref{sec:results_energy}), it qualifies as a superflare and provides an ideal opportunity to investigate the detailed spectral evolution of such an energetic event.}

Figure~\ref{fig:HA1} illustrates the spectral evolution of the H$\alpha$ line profile, which we use as the primary example owing to its high SNR and strong flare response. The right panel shows the corresponding equivalent width excess ($\Delta\mathrm{EW}$), derived using Equation~\ref{eqn:DEW}. This curve reveals a typical flare evolution, with a rapid rise ($\sim$45 minutes) followed by an extended gradual decay ($\sim$1.6 hours) and  a peak excess of $\Delta\mathrm{EW}_{\max}\simeq0.52$~\AA. Given the clear temporal separation of these phases, we focus our analysis on the energy budget and temporal behavior of this well-resolved event.

{\refbf We tracked the same flare evolution in a set of complementary chromospheric diagnostics (Table~\ref{tab:lineparameters}). For each line we compare the reference quiescent spectrum with the profile at the peak of $\Delta\mathrm{EW}$, adopting line-specific integration windows to account for differences in intrinsic width, blending, and SNR. Figure~\ref{fig:CA1} presents the time-resolved profiles and strict $\Delta\mathrm{EW}$ light curves for the Ca~{\sc ii} H\&K resonance lines, which show the strongest flare response after the Balmer series. The remaining diagnostics---the higher-order Balmer, He~{\sc i}, Na~{\sc i}, Mg~{\sc i}, Fe~{\sc ii}, and Ca~{\sc ii} infrared triplet lines---are shown in \ref{sec:lineprofiles}.} Together, these measurements confirm the simultaneous detection of the superflare across multiple chromospheric diagnostics and provide the basis for the energy and timing analyses presented in the following sections.

\begin{figure*}[t]
\centering
\includegraphics[width=180mm]{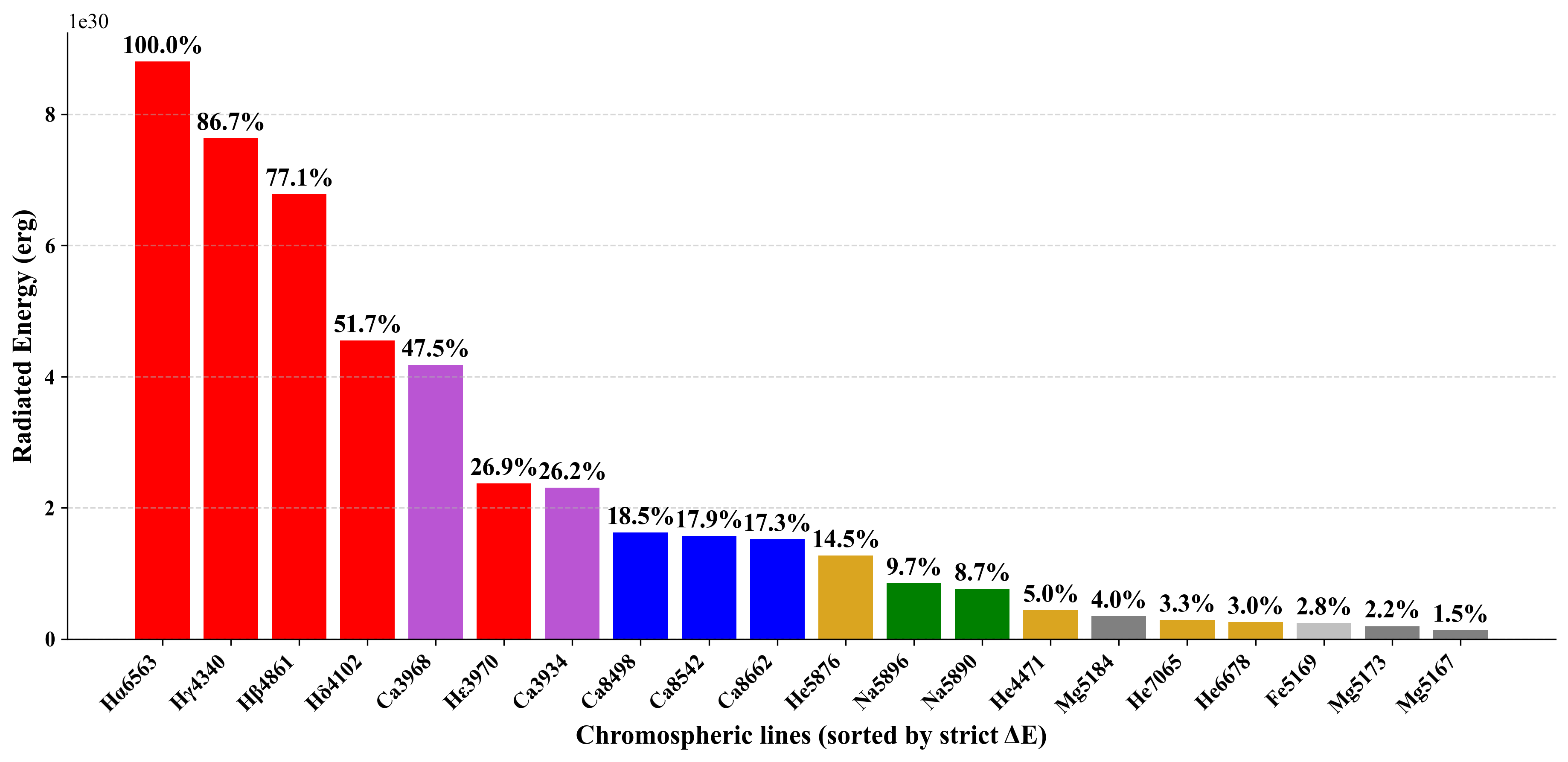}
\caption{Relative radiated energy of chromospheric lines for the 2023 March 14 superflare, arranged in descending order. Energies are normalized to H$\alpha$ (100\%), with other values expressed as percentages relative to H$\alpha$. Colors denote different elements to emphasize the energy distribution by line group.
\label{fig:lines}
}
\end{figure*}

\subsection{Radiated Energy Budget} \label{sec:results_energy} 

Using the methodology detailed in Section~\ref{sec:analysis}, we quantified the radiated energy for all detected chromospheric lines. Focusing on H$\alpha$ as a representative example, we adopted the quiescent continuum flux ($F^q_c = 1.46 \times 10^{-12}$~erg s$^{-1}$ cm$^{-2}$~\AA$^{-1}$; \citealt{2013ApJS..207...15K}) and combined it with the non-negative excess equivalent-width curve, {\refbf $\Delta \mathrm{EW}^{+}(t)$, defined in Section~\ref{sec:analysis}. The total radiated energy was then obtained by integrating the corresponding excess luminosity over the observed time series using the trapezoidal rule (Equation~\ref{eqn:DEtrap})}, yielding $\Delta E_{\mathrm{H}\alpha} \approx 8.8 \times 10^{30}$~erg. The derived parameters and energy estimates for all analyzed lines are compiled in Table~\ref{tab:lineparameters}.

Based on the derived H$\alpha$ radiated energy, we estimated the total bolometric energy of the flare. As detailed in the review by \citet{2024LRSP...21....1K}, the optical $U$-band energy typically accounts for about 10\% of the bolometric flare energy \citep{2015ApJ...809...79O}, while the H$\alpha$ emission corresponds to 4--8\% of the $U$-band energy. This implies that the H$\alpha$ contributes roughly 0.4--0.8\% to the total bolometric energy. Applying this conversion yields a bolometric energy between $1.1 \times 10^{33}$~erg and $2.2 \times 10^{33}$~erg. This estimate firmly places the event in the superflare regime ($>10^{33}$~erg), comparable in scale to the Carrington event observed on the Sun.

{\refbf Among the analyzed chromospheric lines, H$\alpha$ exhibited the largest radiated energy, so we normalized all energies to that of H$\alpha$ (100\%; Figure~\ref{fig:lines}). The next Balmer lines, H$\gamma$ and H$\beta$, reached comparable levels (86.7\% and 77.1\%), followed by H$\delta$ and the Ca~{\sc ii} H line (51.7\% and 47.5\%) and then H$\epsilon$ and the Ca~{\sc ii} K line (26.9\% and 26.2\%). The Ca~{\sc ii} infrared triplet formed a distinct group with moderate, nearly equal contributions (17.3--18.5\% each). Among the helium lines, He~{\sc i} 5876~\AA{} was the strongest at 14.5\%, while the remaining helium and metallic lines (Na~{\sc i}, Mg~{\sc i}, Fe~{\sc ii}) each radiated $\lesssim10\%$ of the H$\alpha$ energy.}

We also examined the Balmer decrement. While the energy emitted in Balmer lines {\refbf in the optically thin case decreases in the order of $\mathrm{H}\alpha > \mathrm{H}\beta > \mathrm{H}\gamma$, flare Balmer decrements are governed by the optical depth and density of the emitting region and can deviate from this ordering} \citep{2003ApJ...597..535H,2013ApJS..207...15K}. {\refbf In our measurements,} the radiated energy in H$\gamma$ exceeds that of H$\beta$ by approximately 12\%. {\refbf Such a reversal can arise physically when the lower-order lines become optically thick, but in our case it may also be affected by the lower SNR and continuum-placement uncertainty at the blue end of our spectra}.

Considering only the chromospheric lines analyzed in the visible and near-infrared (NIR) ranges, we find that the energy radiated at wavelengths longer than 4000~\AA{} is approximately 4.2 times the H$\alpha$ energy. The contribution from wavelengths shorter than 4000~\AA{} is approximately comparable to the H$\alpha$ energy, so that the total radiated energy from all analyzed chromospheric lines amounts to about 5.2 times the H$\alpha$ output. We emphasize that this estimate accounts for line radiation only. Given that the optical continuum can represent a significant fraction of the energy budget \citep[e.g., $\sim$16\% of the bolometric output;][]{2015ApJ...809...79O}, our estimate should therefore be regarded as a lower limit to the total optical/NIR radiative output.

\begin{figure*}[t]
\centering
\includegraphics[width=160mm]{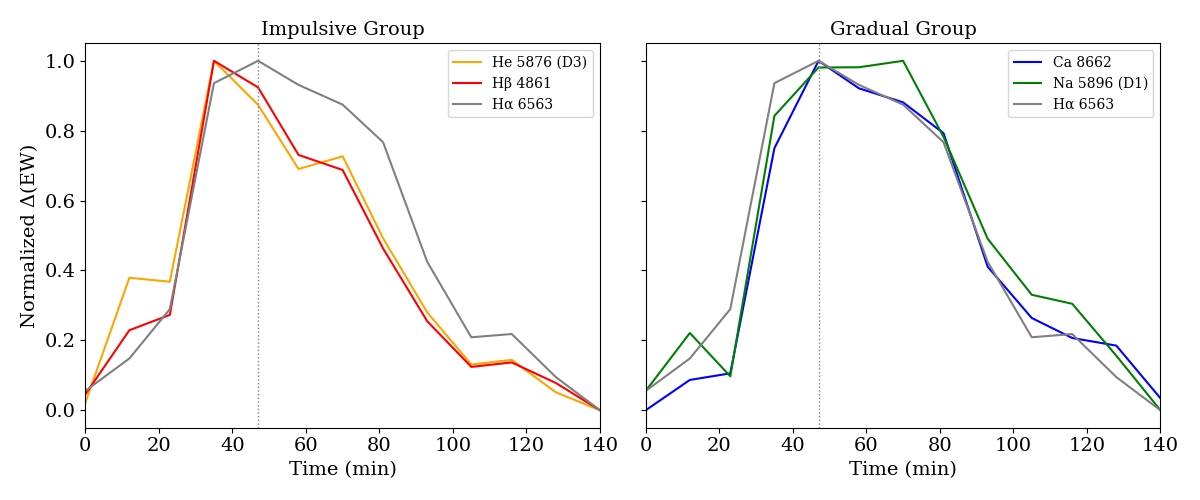}
\caption{Comparison of the $\Delta$EW time evolution for the impulsive (left) and gradual (right) groups. The impulsive group includes Balmer and helium lines, while the gradual group consists of sodium and calcium lines. The total H$\alpha$ profile is included in both panels (gray) for comparison.
\label{fig:r1}}
\end{figure*}

\begin{figure*}[t]
\centering
\includegraphics[width=160mm]{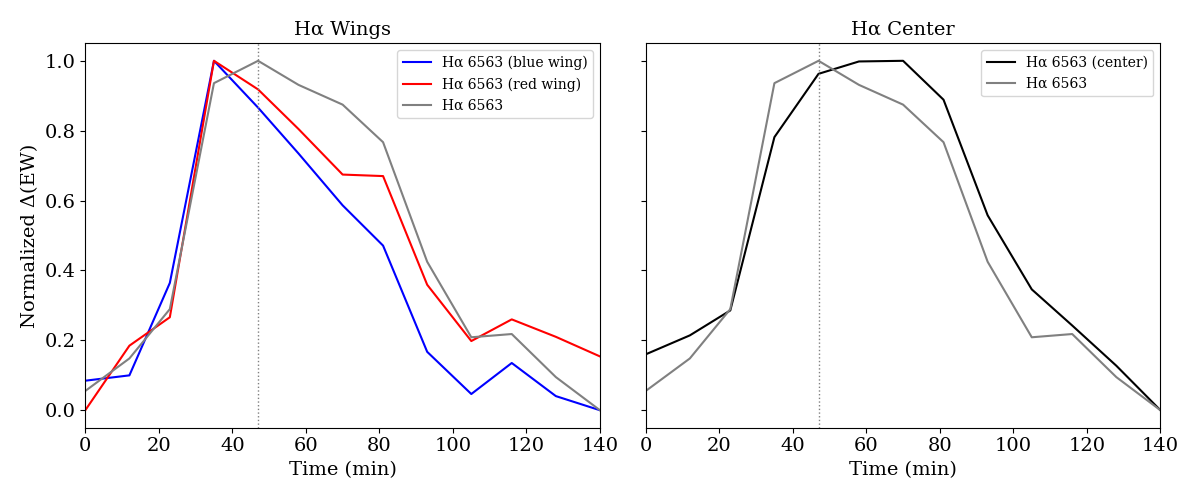}
\caption{Decomposition of the H$\alpha$ profile into wing (left) and center (right) components. The wings trace the impulsive evolution of helium lines, while the center displays a delayed, gradual response relative to the total H$\alpha$ profile (gray) \label{fig:r2}}
\end{figure*}

\subsection{Temporal Evolution of Line Profiles} 
\label{sec:results_time}
The temporal evolution of chromospheric lines during the superflare reveals distinct patterns. As illustrated in Figure~\ref{fig:r1}, the analyzed lines can be broadly categorized into two {\refbf characteristic types of behavior} based on their light curve morphology. The first group, termed the impulsive group, includes neutral helium lines (He~{\sc i}) and higher-order Balmer (e.g., H$\beta$). These lines are characterized by rapid rises to maximum intensity followed by immediate declines. The second group, the gradual group, consists of metallic lines such as the Ca~{\sc ii} infrared triplet and Na~{\sc i}. These lines display delayed peaks relative to the impulsive group and exhibit extended, gradual decay phases.

Notably, the H$\alpha$ line exhibits a behavior that lies on the boundary between these two {\refbf types}. To investigate this hybrid nature, we utilized the high SNR of H$\alpha$ to decompose its profile into three components. This approach is motivated by the fact that the optically thick line core and the optically thin wings probe different atmospheric depths. We defined the line center component by integrating flux over a velocity interval of $\pm20~\mathrm{km\,s^{-1}}$ centered on the line core (corresponding to $\pm0.44$~\AA). Similarly, the blue and red wing components were derived using the same velocity width ($\pm20~\mathrm{km\,s^{-1}}$) but centered at offsets of $-1.2$~\AA\ and $+1.2$~\AA\ from the line core, respectively. Figure~\ref{fig:r2} presents the time evolution of these resolved components. This decomposition disentangles the hybrid behavior of the total H$\alpha$ flux: the H$\alpha$ wings closely trace the impulsive evolution of He~{\sc i}, whereas the line center displays a delayed response and a gradual decay, comparable to the trends observed in metallic lines (e.g., Ca~{\sc ii}).

\begin{figure}[t]
\centering
\includegraphics[width=\linewidth]{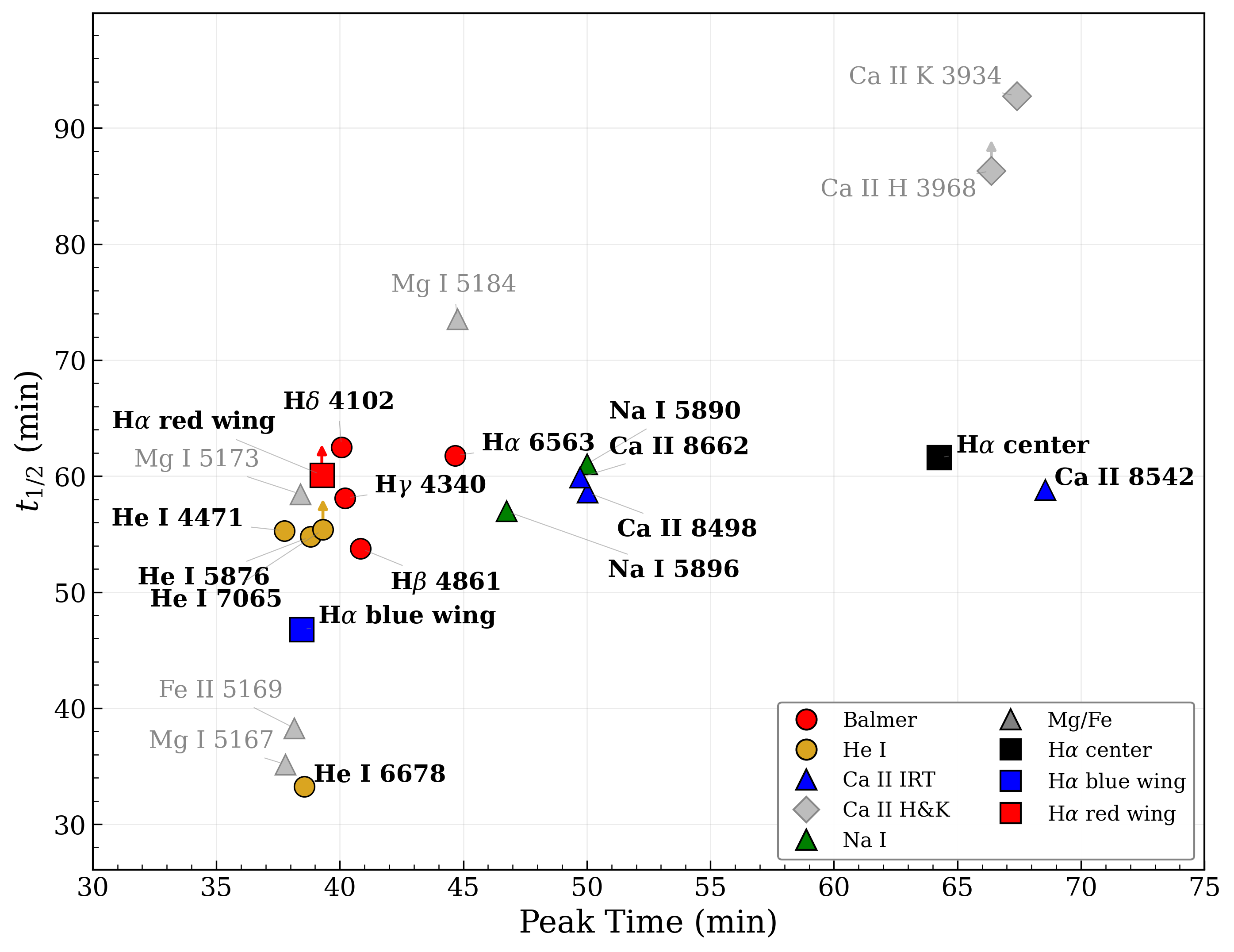}
\caption{{\refbf Peak time versus half-maximum duration $t_{1/2}$ for the chromospheric lines of the 2023 March 14 superflare. The H$\alpha$ center, blue-wing, and red-wing components are shown as squares (black, blue, and red). The other lines are colored by the line group (see legend). Gray symbols mark low-SNR or weak metallic lines (Mg~{\sc i}, Fe~{\sc ii}, and Ca~{\sc ii} H\&K), whose $t_{1/2}$ values are less reliable. Upward arrows denote lower limits on $t_{1/2}$ for lines whose emission had not decayed to half-maximum within the observation window.\label{fig:r3}}}
\end{figure}

{\refbf Figure~\ref{fig:r3} shows the peak time $t_p$ versus the duration $t_{1/2}$ for each line. We take the peak time from a local quadratic fit near the maximum, and $t_{1/2}$ as the full width of the light curve at half-maximum, following \citet{2013ApJS..207...15K}. Upward arrows denote lower limits on $t_{1/2}$ for lines whose emission had not decayed to half-maximum within the observation window. The H$\epsilon$ line is omitted from this diagram because a secondary rise after its peak (visible in Figure~\ref{fig:OT1}) makes its half-maximum width unreliable. The lines separate primarily in peak time: the hydrogen Balmer and He~{\sc i} lines peak earliest ($\sim$37--40~min), whereas the Ca~{\sc ii} resonance and infrared-triplet lines and the Na~{\sc i}~D lines peak $\sim$10--25~min later. The most delayed and longest-lived emission comes from the Ca~{\sc ii} H\&K resonance lines ($t_p\sim$\,66--67~min, $t_{1/2}\gtrsim$\, 86~min). The resolved H$\alpha$ components span this range: the H$\alpha$ center peaks latest ($t_p\sim$\,64~min) with a long $t_{1/2}$, behaving like the gradual metallic lines, while the H$\alpha$ wings peak early ($t_p\sim$\,38--39~min) together with the impulsive Balmer and helium emission. This supports the dual nature of the H$\alpha$ line inferred from the profile decomposition, with the wings tracing the prompt impulsive response and the core tracing a delayed,
gradual evolution.}

\section{Discussion}
\label{sec:discussion}

\subsection{Stratified Chromospheric Response: Evolution of Impulsive and Gradual Emissions}
\label{sec:stratified_response}

The distinct temporal behaviors of the chromospheric lines observed in this study reflect the vertical stratification of the flaring atmosphere. As identified in Section~\ref{sec:results_time}, the emission lines are classified into two morphological groups: an impulsive group (He~{\sc i}, Balmer lines, H$\alpha$ wings) and a gradual group (Ca~{\sc ii}, Na~{\sc i}, H$\alpha$ core). This morphological dichotomy is consistent with seminal spectroscopic observations of M-dwarf flares \citep[e.g.,][]{1991ApJ...378..725H, 2003ApJ...597..535H, 2013ApJS..207...15K}, providing crucial observational constraints on the relative depths of energy deposition.

These temporal trends are well interpreted within the framework of the Neupert effect \citep{1968ApJ...153L..59N}, which describes the causal link between particle acceleration and atmospheric heating. In solar flares, impulsive hard X-ray emission correlates with the time derivative of gradual soft X-ray emission, reflecting the conversion of kinetic energy into a thermal response. Our observations exhibit a similar phenomenology: the sharp peaks of the He~{\sc i}, Balmer lines and H$\alpha$ wings trace instantaneous energy deposition, while the delayed metallic lines and H$\alpha$ core correspond to the cumulative thermal response of the cooling chromosphere. Although a quantitative test of this effect is limited by our 12-minute cadence, the observed offsets between impulsive and gradual components underscore the complex heating dynamics of the flaring atmosphere.

This thermal phase is most clearly characterized by the delayed peak and prolonged decay of the Ca~{\sc ii} H\&K lines \citep{1991ApJ...378..725H, 2013ApJS..207...15K}. \citet{2013ApJS..207...15K} interpreted this evolution as a signature of heated plasma accumulating energy from the impulsive phase. We confirm that this behavior extends beyond resonance lines; the Ca~{\sc ii} infrared triplet and Na~{\sc i} lines exhibit a similar delayed response, serving as effective tracers of the thermally dominated phase. 

The hybrid behavior of the H$\alpha$ line profile provides more direct evidence for such a depth-dependent atmospheric response. In our observations, the H$\alpha$ wings exhibit sharp, impulsive variations consistent with Stark (pressure) broadening induced by non-thermal energy deposition in the dense lower chromosphere \citep{2006PASP..118..227P}. Radiative-hydrodynamic simulations confirm that these broad wing components are a direct consequence of significantly increased electron density in deeper layers \citep{2020PASJ...72...68N}. In contrast, the H$\alpha$ line center remains less sensitive to the initial impulsive heating and exhibits a gradual evolution consistent with the behavior of metallic lines.

\begin{table}
\centering
\caption{Chromospheric line energy ratios normalized to H$\alpha$ for an M7.7-class solar flare, AD Leo moderate flares, and our superflare.
\label{tab:comparison}}
\begin{tabular}{l c c c}
\hline\hline
Line & Solar$^a$ & AD Leo (H03)$^b$ & This Study \\
\hline
H$\beta$                      & 52.5 & 62--103 & 77.1 \\
Ca\,\textsc{ii} K (3934\,\AA) & 81.2 & \ldots$^c$ & 26.2 \\
Ca\,\textsc{ii} H (3968\,\AA) & 87.3 & \ldots$^c$ & 47.5 \\
Ca\,\textsc{ii} 8662\,\AA     & 40.8 & 19--53 & $\sim$17 \\
Ca\,\textsc{ii} 8498\,\AA     & 27.9 & 8--42  & $\sim$18 \\
He\,\textsc{i} 5876\,\AA      & \ldots$^d$ & 14--35 & $\sim$15 \\
\hline
\end{tabular}

\smallskip
\begin{minipage}{0.48\textwidth}
\small
\textit{Note.}---Radiated energy of each line normalized to
H$\alpha$ = 100\%. \\
$^a$ M7.7-class solar flare (1993 March 6);
computed from Johns-Krull et al.\ (1997), Table~4. \\
$^b$ Hawley et al.\ (2003), Table~6 (sum of impulsive and
gradual phases); ranges span flares~3, 6, 7, and 8. \\
$^c$ Ca\,\textsc{ii}\,K not measured in H03;
Ca\,\textsc{ii}\,H unreliable due to H$\epsilon$ blending
and low blue-end SNR. \\
$^d$ He\,\textsc{i}\,5876\,\AA\ not included in the
Johns-Krull et al.\ (1997) energy budget.
\end{minipage}
\end{table}

\subsection{Comparison with Previous AD~Leo Flare Studies\label{sec:comparison_adleo}}
{\refbf AD~Leo flares have been spectroscopically characterized over a wide energy range, from $\sim$$10^{29}$\,erg non-white-light events \citep{2006A&A...452..987C} to the $>$$10^{34}$\,erg giant flare of \citet{1991ApJ...378..725H}. With $E_{\mathrm{bol}}\sim10^{33}$\,erg, our event bridges these energy levels. 

\citet{2003ApJ...597..535H} analyzed eight flares on AD~Leo using simultaneous optical ($R\sim55{,}000$) and UV spectroscopy, providing the most detailed chromospheric energy budget available for comparison. Table~\ref{tab:comparison} compares the line energy ratios normalized to H$\alpha$ between their four well-characterized flares and our superflare. The Ca\,{\sc ii} infrared triplet provides the most robust basis for comparison, as both datasets have reliable measurements in this red spectral region. The Ca\,{\sc ii} 8498\,\AA\ and 8662\,\AA\ lines contribute $\sim$18\% and $\sim$17\% of the H$\alpha$ energy in our superflare, compared to 8--42\% and 19--53\% across the \citet{2003ApJ...597..535H} flares, respectively. The He\,{\sc i} 5876\,\AA\ line shows a similar pattern ($\sim$15\% vs.\ 14--35\%). Our superflare therefore falls within the range spanned by the \citet{2003ApJ...597..535H} flares, indicating broadly consistent relative line contributions despite its much higher total energy.}

\subsection{Comparison with a Solar Flare\label{sec:results3}}

A comparison with the M7.7-class solar flare analyzed by \citet{1997ApJS..112..221J} provides critical context for interpreting the physical scale and atmospheric response of the AD~Leo superflare. As solar flares serve as the primary laboratory for understanding flare physics, comparing stellar data to a well-resolved solar analog is key to determining how these physical mechanisms scale to much higher energy levels. \citet{1997ApJS..112..221J} provides a unique reference with simultaneous high-resolution (R $\sim 48,000$) spectra covering 3800--9000 \AA. This rare time-resolved observation of a large solar flare serves as a valuable standard for detailed multi-line comparisons. While the solar flare released approximately $4.5 \times 10^{29}$ erg in the H$\alpha$ line (bolometric energy $\sim 10^{31}$ erg), the AD~Leo event is nearly two orders of magnitude more energetic. Despite this large difference in total energy, both flares share the feature of H$\alpha$ remaining the dominant chromospheric diagnostic.

\begin{figure}
\centering
\includegraphics[width=80mm]{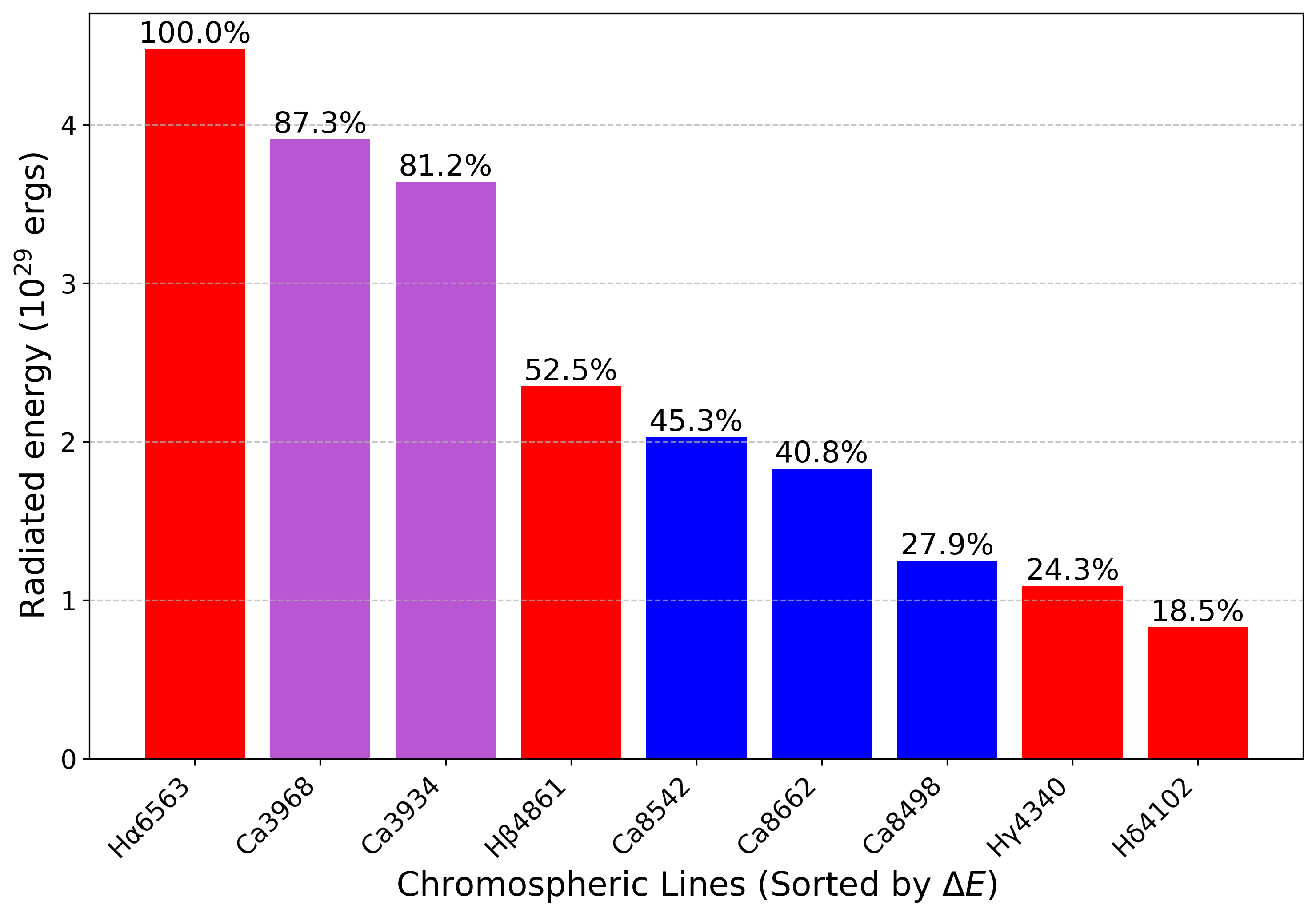}
\caption{Relative energy distribution of chromospheric lines in an M7.7-class solar flare, arranged in descending order. Energies are normalized to H$\alpha$ (100\%), with other values expressed as percentages relative to H$\alpha$. Constructed from the data of \citet{1997ApJS..112..221J}.
\label{fig:r4}}
\end{figure}

A key distinction, however, lies in the relative energy distribution among the chromospheric lines. In the solar flare, the Ca\,{\sc ii} H and K lines radiate energy at levels highly comparable to H$\alpha$, accounting for 87.3\% and 81.2\% of its flux, respectively (see Figure \ref{fig:r4}). In contrast, the AD~Leo superflare exhibits a much weaker contribution  relative to H$\alpha$, with the Ca\,{\sc ii} H and K lines contributing only 47.5\% and 26.2\%, respectively (see Figure~\ref{fig:lines}). This discrepancy suggests a fundamental difference in the chromospheric density and optical depth between solar and dMe stellar flares. In the solar atmosphere, where densities are relatively low, flare energy is primarily released through the H$\alpha$ and Ca\,{\sc ii} resonance lines. However, in dMe flares, the higher chromospheric densities lead to larger optical depths in these lower-order transitions, effectively shifting the cooling contribution toward the Balmer continuum and optically thinner, higher-order Balmer transitions \citep{1997ApJS..112..221J,2013ApJS..207...15K}.

Another informative comparison focuses on the temporal evolution of Ca\,{\sc ii} emission. In the solar flare, the Ca\,{\sc ii} infrared triplet lines were found to decline more slowly than the resonance lines (see Fig. 5 of \citealt{1997ApJS..112..221J}). Our AD~Leo superflare data exhibit qualitatively similar behavior, with the entire triplet—most notably the 8662 \AA\ line—showing a noticeably more prolonged decay phase compared to the impulsive Balmer emission. However, the degree of this gradual-phase dominance differs between the two flares. In the solar event, the Ca\,{\sc ii} decay is extended but modest, reflecting the relatively short heating timescale of the M7.7 flare. In the AD~Leo superflare, the same qualitative trend is present but stretched over a much longer duration, consistent with the substantially larger energy scale and prolonged energy release of the stellar flare.

Overall, these comparisons indicate that chromospheric line behavior in stellar flares cannot be directly inferred from solar analogs. Differences in chromospheric density, optical depth, and energy transport processes—together with the much larger flare energies on active M-dwarfs—play a critical role in determining the observed line emission. Accounting for these factors is essential when applying solar flare models to stellar flare environments.

\section{Summary\label{sec:summary}}
This study presents high-resolution, time-resolved spectroscopy of the active M-dwarf AD~Leo, obtained using the 1.8~m telescope and BOES spectrograph at BOAO. The observations captured the complete evolution of a superflare event on 2023 March 14. By analyzing variations in the equivalent width of chromospheric lines, we quantified the radiated energy budget and investigated the vertical stratification of the flaring atmosphere.

{\refbf The radiated energy of the superflare, derived from the H$\alpha$ line, was $\sim 8.8 \times 10^{30}$~erg, implying a bolometric energy of $\sim10^{33}$~erg. The Balmer series dominated the chromospheric line-energy budget. The Ca\,{\sc ii} H and K resonance lines were the next strongest contributors, radiating $\sim$48\% and $\sim$26\% of the H$\alpha$ energy, followed by the Ca\,{\sc ii} infrared triplet ($\sim$17--18\% each) and the He\,{\sc i} 5876~\AA\ line ($\sim$15\%). The remaining metallic lines each contributed below $\sim$10\%. These relative strengths provide an empirical baseline for estimating energy budgets in future studies of energetic M-dwarf flares.}

Temporal analysis revealed that the chromospheric lines fall into two {\refbf broad} morphological groups. The impulsive group, characterized by rapid rises and declines, consists of the Balmer and He~{\sc i} lines. The gradual group, exhibiting delayed peaks and extended decay phases, includes metallic lines such as the Ca~{\sc ii} infrared triplet and Na~{\sc i}. The H$\alpha$ line exhibited a hybrid behavior; its wings traced the impulsive phase, consistent with Stark broadening in the lower chromosphere, while its core evolved gradually. These behaviors are consistent with the Neupert effect and reflect the vertical stratification of the energy deposition.

A comparison with an M7.7-class solar flare {\refbf reveals different energy partitioning. Although H$\alpha$ dominates the chromospheric line budget in both cases, the relative strength of the Ca\,{\sc ii} resonance lines is much lower in the AD~Leo superflare than in the solar event.}This contrast indicates that solar flare models cannot be linearly extrapolated to M-dwarf superflares without accounting for the different density and temperature structure of the M-dwarf chromosphere.

This study is limited by the absence of simultaneous photometry, which introduces uncertainty in absolute energy estimates. Furthermore, the temporal resolution was constrained by the exposure times required for sufficient SNRs. Future investigations utilizing high-cadence spectroscopy and simultaneous multi-band photometry are essential to resolve rapid continuum variations and flare substructures in active stellar atmospheres.


\acknowledgments
We thank the anonymous referee for constructive comments that improved the manuscript. S.-W.C. acknowledges support from the National Research Foundation of Korea (NRF) through grants RS-2026-25489059 and RS-2026-25490019, funded by the Korean government (MSIT), and through the Basic Science Research Program (RS-2023-00245013), funded by the Ministry of Education. J.C. acknowledges support from the NRF under grant RS-2023-00208117, and K.-S.L. acknowledges support from the NRF under grant RS-2025-23523356. J.K. and E.-K.L. acknowledge support by the Korea Astronomy and Space Science Institute under the R\&D program of the Korean government (MSIT; No. 2026-1-830-05).

This paper was fully based on observations obtained at the Bohyunsan Optical Astronomy Observatory (BOAO), which is operated by the Korea Astronomy and Space Science Institute (KASI). We thank the staff of the BOAO, including the telescope operators, for their dedicated support and technical assistance during the observing runs, which were essential in obtaining the data for this study. 


\appendix

\section{Additional Line Profiles}\label{sec:lineprofiles}
{\refbf Here we present the line-profile changes and $\Delta$EW light curves for the chromospheric diagnostics not shown in the main text. As in Figure~\ref{fig:CA1}, the left panels of each figure compare the quiescent spectrum with the profile at the peak of $\Delta$EW, and the right panels show the corresponding $\Delta$EW light curves. The integration windows adopted for each line are listed in Table~\ref{tab:lineparameters}. These figures confirm that the superflare is detected simultaneously across all analyzed chromospheric lines.}

\begin{figure}[!t]
\centering
\includegraphics[width=\columnwidth, keepaspectratio]{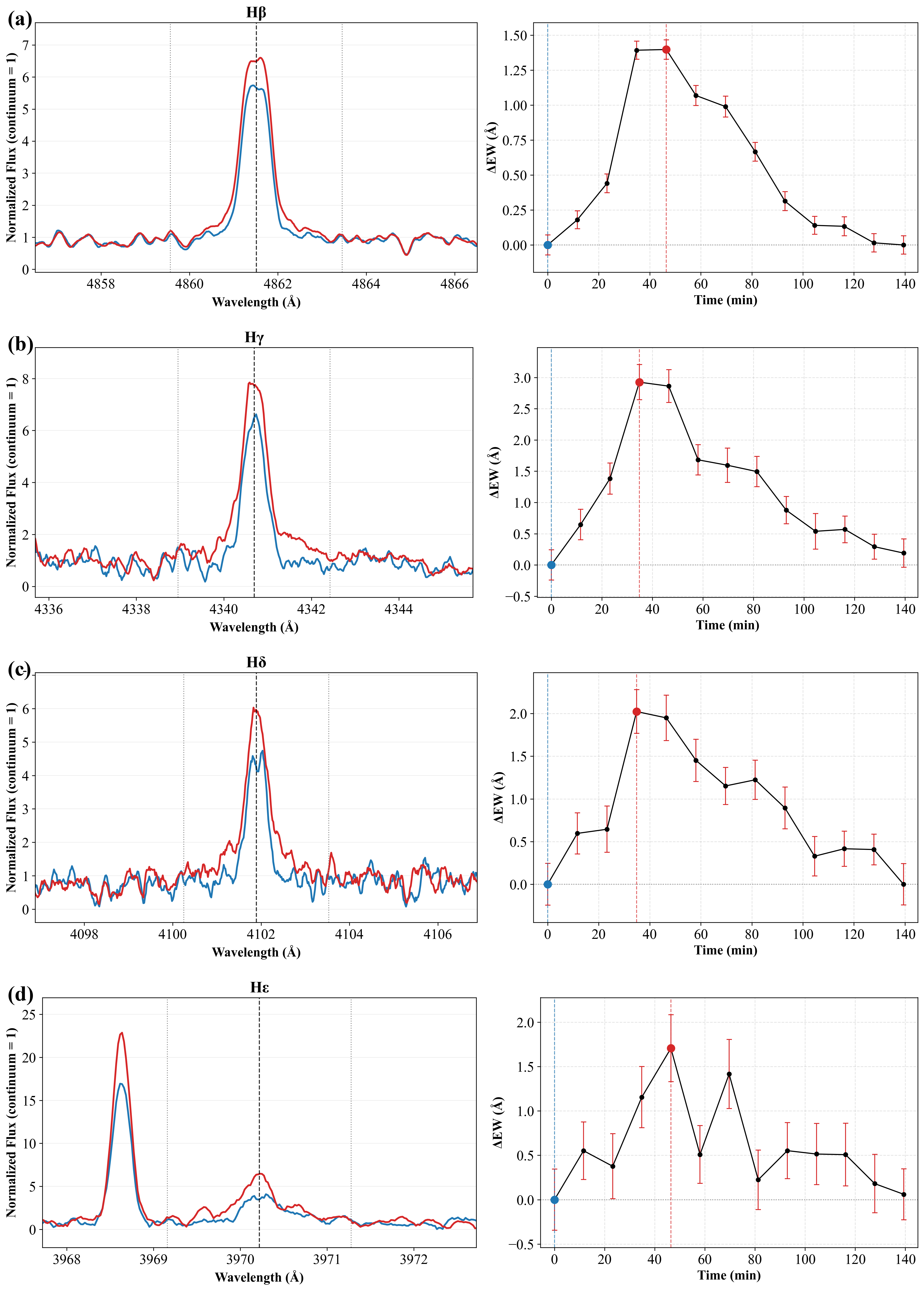}
\caption{Same as Figure~\ref{fig:CA1}, but for the higher-order Balmer lines: (a) H$\beta$ 4861~\AA, (b) H$\gamma$ 4340~\AA, (c) H$\delta$ 4102~\AA, and (d) H$\epsilon$ 3970~\AA.
}
\label{fig:OT1}
\end{figure}

\begin{figure}[!t]
\centering
\includegraphics[width=\columnwidth, keepaspectratio]{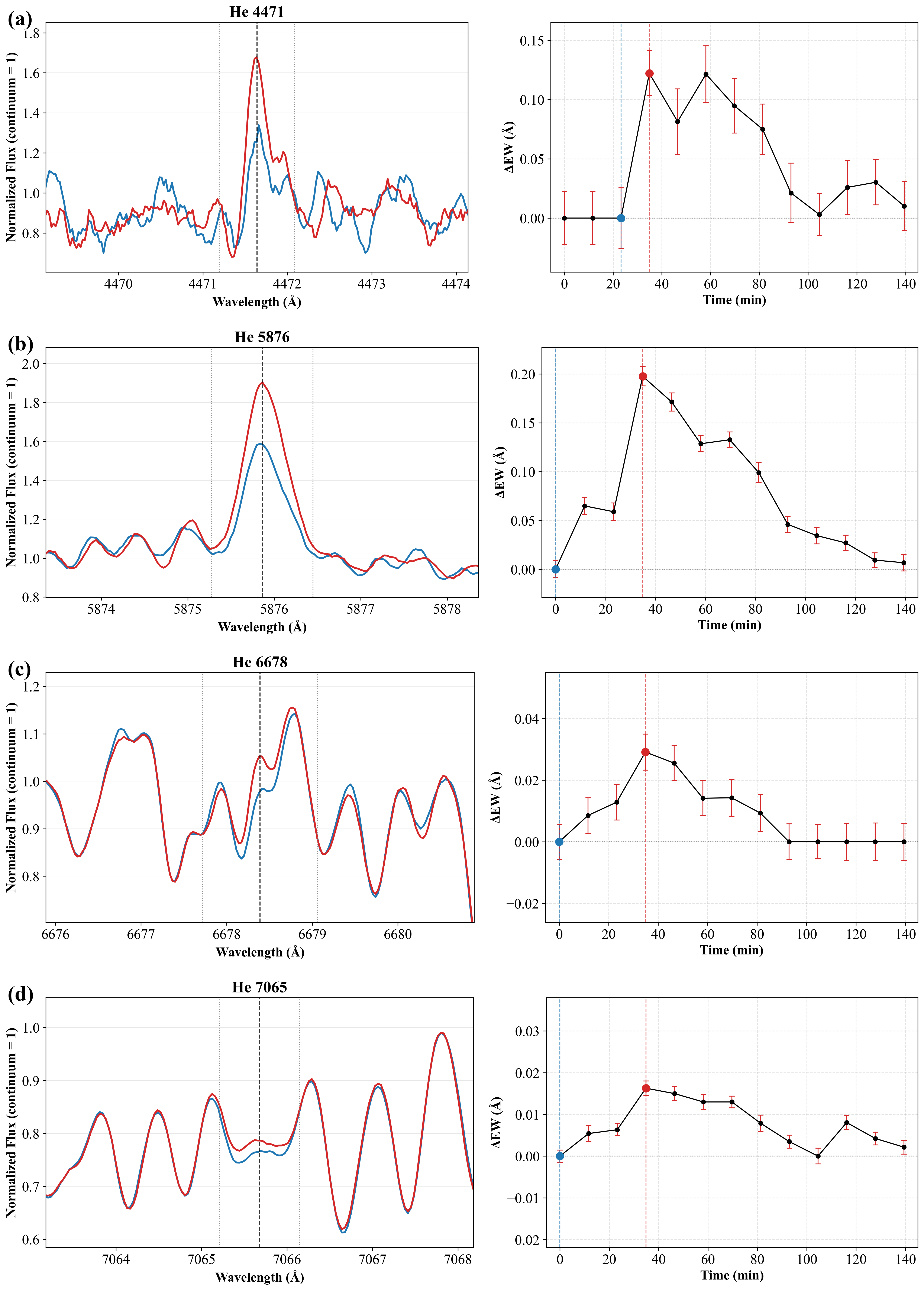}
\caption{Same as Figure~\ref{fig:CA1}, but for the He~{\sc i} lines: (a) He~{\sc i} 4471~\AA, (b) He~{\sc i} 5876~\AA, (c) He~{\sc i} 6678~\AA, and (d) He~{\sc i} 7065~\AA.
}
\label{fig:OT2}
\end{figure}

\begin{figure}[!t]
\centering
\includegraphics[width=\columnwidth, keepaspectratio]{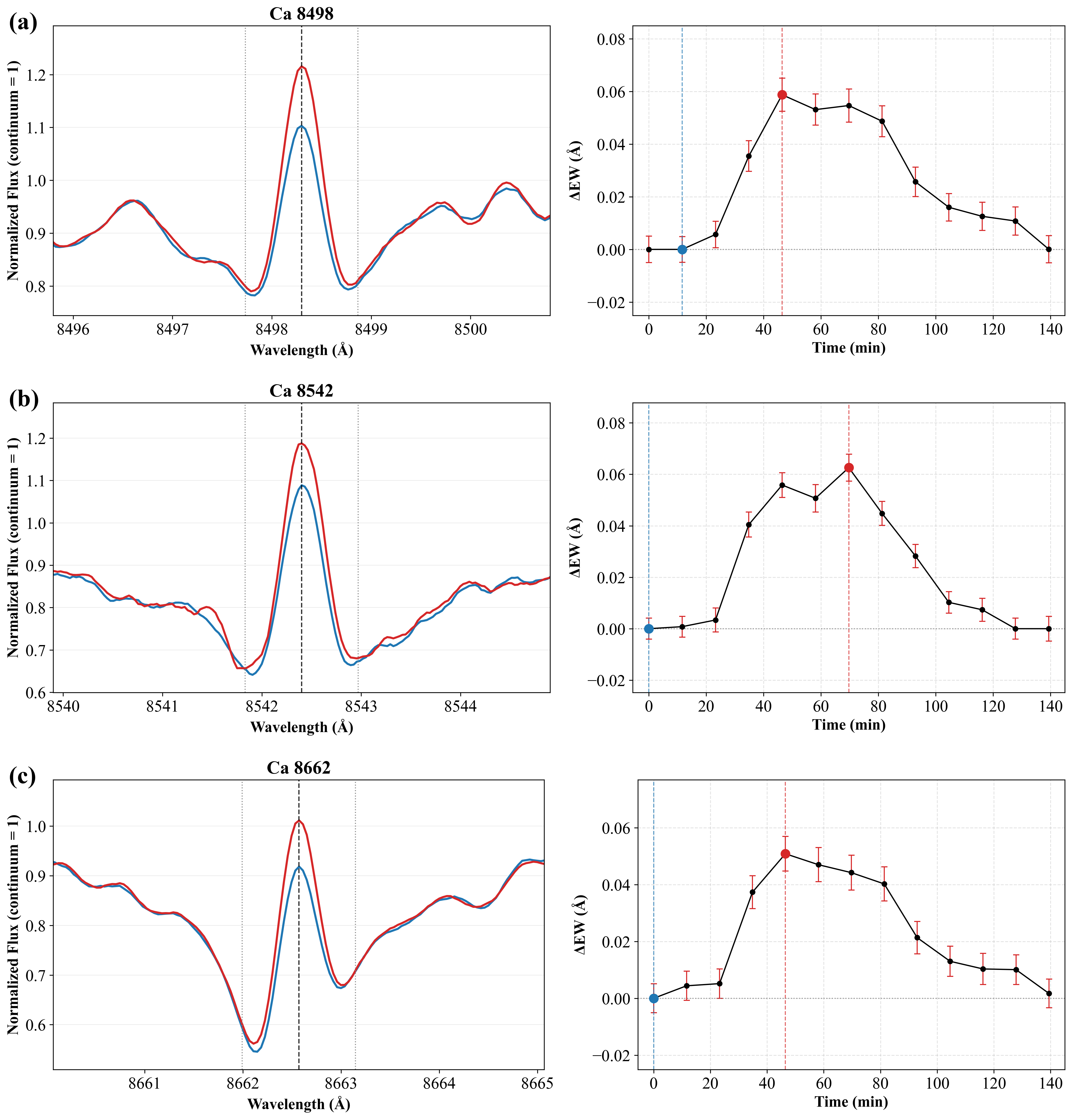}
\caption{Same as Figure~\ref{fig:CA1}, but for the Ca~{\sc ii} infrared triplet lines: (a) Ca~{\sc ii} 8498~\AA, (b) Ca~{\sc ii} 8542~\AA, and (c) Ca~{\sc ii} 8662~\AA.
}
\label{fig:OT3}
\end{figure}

\begin{figure}[!t]
\centering
\includegraphics[width=\columnwidth, keepaspectratio]{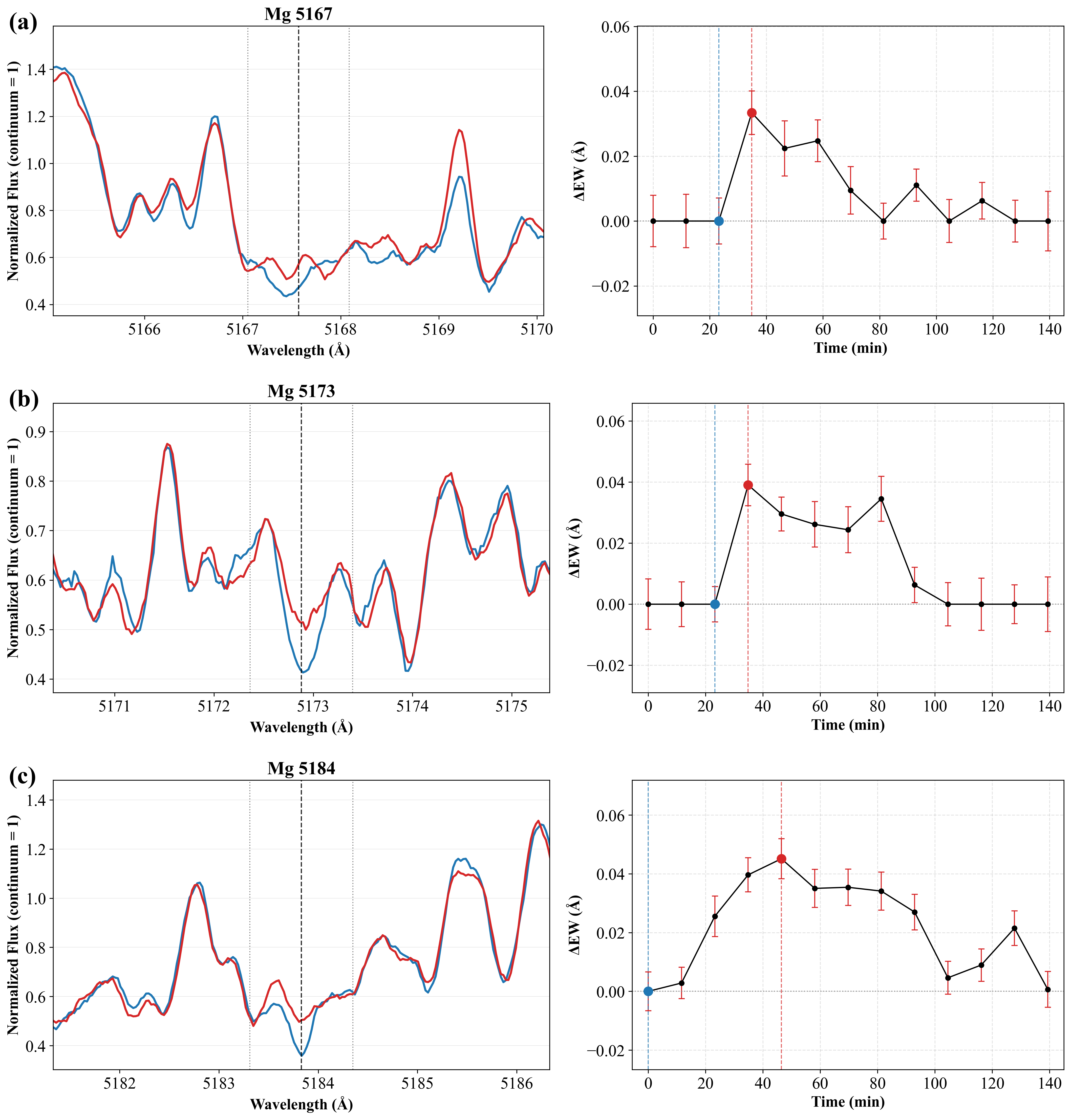}
\caption{Same as Figure~\ref{fig:CA1}, but for the Mg~{\sc i} triplet lines: (a) Mg~{\sc i} 5167~\AA, (b) Mg~{\sc i} 5173~\AA, and (c) Mg~{\sc i} 5184~\AA.
}
\label{fig:OT4}
\end{figure}

\begin{figure}[!t]
\centering
\includegraphics[width=\columnwidth, keepaspectratio]{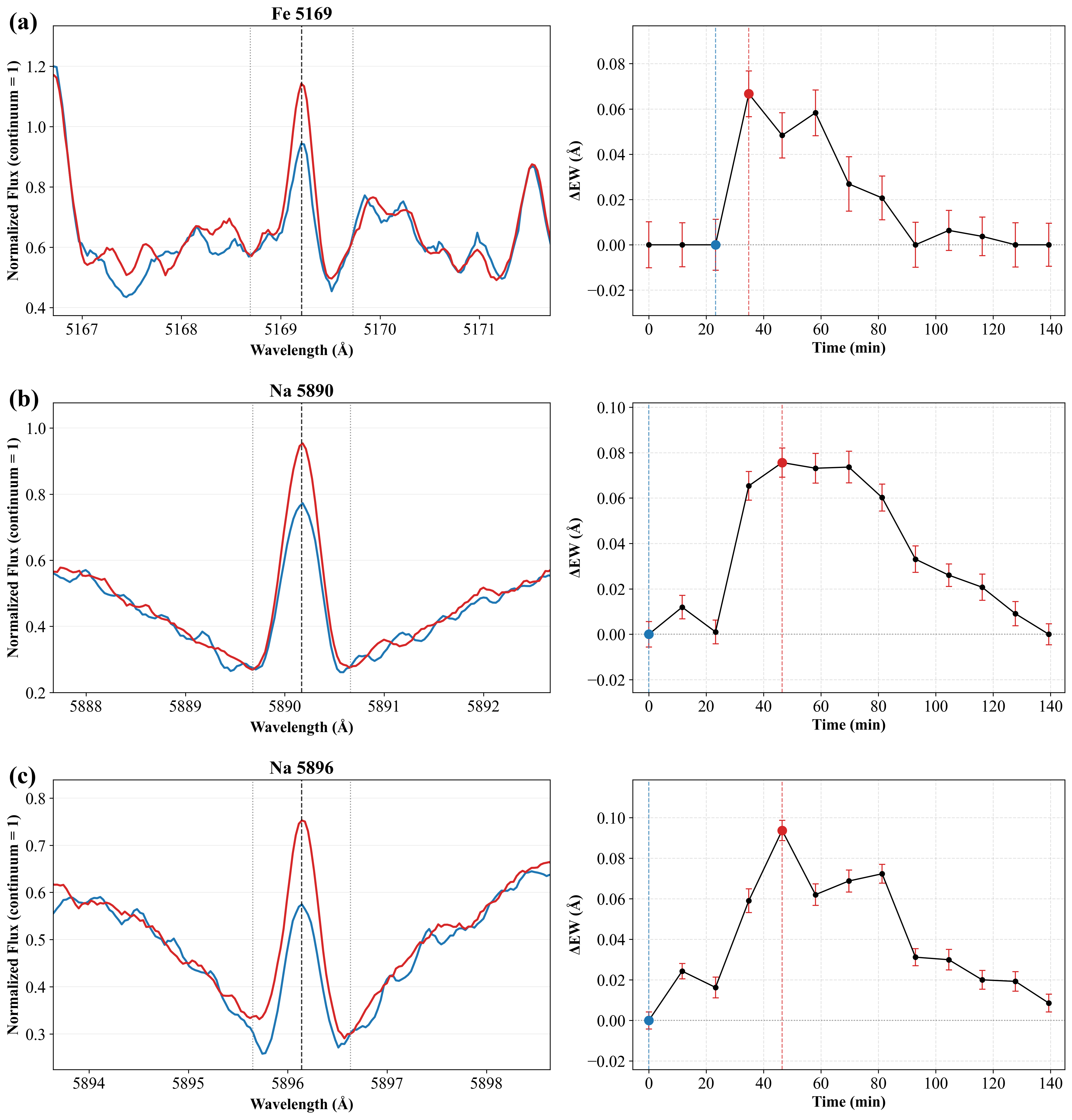}
\caption{Same as Figure~\ref{fig:CA1}, but for the Fe~{\sc ii} and Na~{\sc i} lines: (a) Fe~{\sc ii} 5169~\AA, (b) Na~{\sc i} D2 5890~\AA, and (c) Na~{\sc i} D1 5896~\AA.
}
\label{fig:OT5}
\end{figure}

\bibliography{ADLeo_BOES_superflare}

\end{document}